\documentclass[aps,superscriptaddress,showpacs,floatfix,pdflatex]{revtex4}
\usepackage{graphicx}
\usepackage{longtable}
\usepackage{epsfig}
\usepackage{epstopdf}
\usepackage{CJK}
\usepackage{float}
\ifx\pdfoutput\undefined
\usepackage[dvips]{graphicx}
\usepackage[dvipdfm,
  pdfstartview=FitH,
     CJKbookmarks=true,
          bookmarksnumbered=true,
          bookmarksopen=true,
         colorlinks=true,
          pdfborder=002,
          citecolor=blue%
           ]{hyperref}
\AtBeginDvi{} 
\else
\usepackage[
          pdfstartview=FitH,
          CJKbookmarks=true,
          bookmarksnumbered=true,
           bookmarksopen=ture,
           colorlinks=true,
           citecolor=blue
           ]{hyperref}

\usepackage{txfonts}
\usepackage{booktabs}

\begin{document}

\title{{An exact solution} of spherical mean-field plus orbit-dependent\\
non-separable pairing model
with two non-degenerate $j$-orbits}

\author{Feng Pan\footnote{The corresponding author's e-mail: daipan@dlut.edu.cn}}
\affiliation{Department of Physics, Liaoning Normal University,
Dalian 116029, China}\affiliation{Department of Physics and
Astronomy, Louisiana State University, Baton Rouge, LA 70803-4001,
USA}

\author{Shuli Yuan}
\affiliation{Department of Physics, Liaoning Normal University,
Dalian 116029, China}

\author{Yingwen He}
\affiliation{Department of Physics, Liaoning Normal University,
Dalian 116029, China}

\author{Yunfeng Zhang}
\affiliation{Department of Physics, Liaoning Normal University,
Dalian 116029, China}

\author{Siyu Yang}
\affiliation{Department of Physics, Liaoning Normal University,
Dalian 116029, China}

\author{J. P. Draayer}
\affiliation{Department of Physics and Astronomy, Louisiana State
University, Baton Rouge, LA 70803-4001, USA}

\begin{abstract}
An exact solution of nuclear spherical mean-field plus orbit-dependent non-separable
 pairing model with two non-degenerate $j$-orbits is presented.
The extended one-variable Heine-Stieltjes polynomials associated to the Bethe ansatz
equations of the solution are determined, of which the sets of the zeros give
the solution of the model, and can be determined relatively easily.
A comparison of the solution to that of the standard pairing interaction with constant
interaction strength among pairs in any orbit is made.
It is shown that the overlaps of eigenstates of the model with
those of the standard pairing model are {always large,} especially
for the ground and the first excited state. However,
the quantum phase crossover in the non-separable pairing model
{cannot} be accounted for by the standard pairing interaction.

\end{abstract}


\maketitle
\noindent {\bf 1. Introduction:}~ Pairing correlations seem evident
in various quantum many-body systems.~\cite{bar,ran,coop,gomes}.
It has been shown {that} pairing interactions
are key to {elucidating} ground state and low-energy
spectroscopic properties of nuclei~\cite{Belyaev,Ring,Hasegawa}.
Though the Bardeen-Cooper-Schrieffer (BCS)~\cite{bar}
and the  Hartree-Fock-Bogolyubov (HFB) approximations
provide simple and clear pictures~\cite{Belyaev,PN,ma},
tremendous efforts have been made in finding
{exact} solutions to the problem~\cite{dans,cov,bi,zeng,mol,Volya}.
It is known that spherical or deformed mean-field
plus the standard (equal strength) pairing interaction can be solved
exactly by using the Gaudin-Richardson method~\cite{gau,Ri,duk}, which
can now be solved more easily by using the
extended Heine-Stieltjes polynomial approach~\cite{pan0,guan1, guan2,qi}.
The separable pairing
problem was studied in \cite{pan3},
in which the single-particle energies are all degenerate.
The separable pairing interaction with two non-degenerate levels
was analyzed in \cite{ba}, of which solution with multi non-degenerate levels
of  a special case was given in \cite{Rom,claeys,pan4}, while the general
case has been analyzed in \cite{pan5}.
In this work, it will be shown that the orbit-dependent
non-separable pairing interaction among valence nucleons
over two non-degenerate orbits can  also be solved analytically.

\vskip .3cm
\noindent {\bf 2. The model and exact solution:}~
The Hamiltonian of a spherical mean-field plus orbit-dependent
non-separable pairing model (NSPM) with two non-degenerate $j$-orbits
{can} be written as
\begin{eqnarray}\label{H}
\hat{H} =\sum_{t}^{p}\epsilon_{{t}}\,\hat{N}_{j_{t}}+\hat{H}_{\rm P}=
\sum_{t}^{p}\epsilon_{{t}}\,\hat{N}_{j_{t}}-\sum_{1\leq t,t'\leq p}g_{{t},{t'}}\,S^{+}_{j_{t}}S^{-}_{j_{t'}},
\end{eqnarray}
where $p=2$ is the total number of orbits {considered above a closed or sub-closed shell,}
$\{\epsilon_{{t}}\}$ ($t=1,~2$) is  single-particle energies
generated from a mean-field theory with $\epsilon_{{1}}\neq \epsilon_{{2}}$,
$\hat{N}_{j}=\sum_{m}a^{\dagger}_{jm}a_{jm}$
and $S_{j}^{+}=\sum_{m>0}(-1)^{j-m}a^{\dagger}_{jm}a^{\dagger}_{j-m}$,
in which $a^{\dagger}_{jm}$ ($a_{jm}$)
is the creation (annihilation) operator
for a nucleon with angular momentum quantum number $j$ {with projection} $m$,
and $\{g_{{t},{t'}}\}$ ($t,t'=1,2$) are
the non-separable pairing interaction
parameters, which are all assumed to be real and must be symmetric with $g_{{1},{2}}=g_{{2},{1}}$.

The set of local operators $\{{S}^{-}_{j_{t}},~{S}^{+}_{j_{t}},~
\hat{N}_{j_{t}}\}$ ($t=1,2$), where $S^{-}_{j_{t}}=(S^{+}_{j_{t}})^{\dag}$,
generate two copies of {an SU(2) algebra
that satisfies} the commutation relations
$
[\hat{N}_{j_{t}}/2,~{S}^{-}_{j_{t'}}]=-\delta_{tt'}{S}^{-}_{j_{t}},~
[\hat{N}_{j_{t}}/2,~{S}^{+}_{j_{t'}}]=\delta_{tt'}{S}^{+}_{j_{t}},~
[{S}^{+}_{j_{t}},~{S}^{-}_{j_{t'}}]=2\delta_{tt'}S^{0}_{j_{t}},
$
where $S^{0}_{j_{t}}=(\hat{N}_{j_{t}}-\Omega_{{t}})/2$ with $\Omega_{{t}}=j_{t}+1/2$.
As adopted in the Gaudin-Richardson approach ~\cite{gau,Ri,duk} for the standard pairing model (SPM), let
\begin{equation}\label{2}
S^{+}(x)=\sum_{t}^{2}{1\over{2\epsilon_{{t}}-x}}S_{j_{t}}^{+},
\end{equation}
where $x$ is the spectral parameter to be determined.
According to the commutation relations of the generators of the
two copies of {the} SU(2) algebra, we have
\begin{equation}\label{3}
[\sum_{t}\epsilon_{{t}}\hat{N}_{j_{t}},~S^{+}(x)]=
\sum_{t}{2\epsilon_{{t}}\over{2\epsilon_{{t}}-x}}S^{+}_{j_{t}}=
 S^{+}+x\,S^{+}(x),
\end{equation}
where $S^{+}=\sum_{t} S^{+}_{j_{t}}$, and
\begin{equation}\label{4}
[\hat{H}_{\rm P},~S^{+}(x)]=
\sum_{t',t}g_{{t'}, {t}}S^{+}_{j_{t'}}{2S_{j_{t}}^{0}\over{2\epsilon_{{t}}-x}},
\end{equation}
\begin{equation}\label{5}
[[\hat{H}_{\rm P},~S^{+}(x)], ~S^{+}(y)]=
2\sum_{t',t}g_{{t'},{t}}{1\over{(2\epsilon_{{t}}-x)
(2\epsilon_{{t}}-y)}}S^{+}_{j_{t'}}S_{j_{t}}^{+}.
\end{equation}

The $k$-pair eigenvectors of (\ref{H}) {can} be still written
as the Gaudin-Richardson form with

\begin{equation}\label{6}
\vert \zeta,~k;JM\rangle=\prod_{\rho}^{k}S^{+}(x^{(\zeta)}_{\rho})
\vert JM\rangle,
\end{equation}
where $\zeta$ labels the $\zeta$-th set of solution $\{x^{(\zeta)}_{1},\cdots,x^{(\zeta)}_{k}\}$.
If the seniority number of the $t$-th orbit is $\nu_{{t}}$,
the pairing vacuum states of these two orbits
are denoted as $\vert \nu_{{t}}\eta_{{t}}J_{t}M_{t}\rangle$ satisfying
$S^{-}_{j_{t}}\vert \nu_{j_{t}}\eta_{{t}}J_{t}M_{t}\rangle=0$, where
$J_{{t}}$ and $M_{{{t}}}$ are the angular momentum
quantum number  and that of its third component, respectively,
and
$\eta_{{t}}$ is the multiplicity label needed to distinguish
different possible ways of $\nu_{{t}}$ particles
coupled to the angular momentum $J_{{t}}$.
Thus, a pairing vacuum state of a two $j$-orbit
system with the total seniority number $\nu=\nu_{{1}}+\nu_{{2}}$
and the total angular momentum $J$  can be expressed as $\vert JM\rangle\equiv\vert \nu_{{1}}\eta_{{1}},
\nu_{{2}}\eta_{{2}};(J_{{{1}}}\otimes J_{{{2}}})JM\rangle$.
Thus, $\vert JM\rangle$ satisfies  $S^{-}_{j_{t}}\vert JM\rangle=0$ for $t=1,2$,
which is used in (\ref{6}).

To solve the eigen-equation of (\ref{H}) {with} ansatz (\ref{6}),
one {can} calculate commutators of $\hat{H}$ with the pairing operators
$S^{+}(x^{(\zeta)}_{\rho})$ as was done
in Richardson's work on solving the SPM~\cite{Ri,duk}.
Since (\ref{H}) only contains one- and two-body interaction terms,
the $q$-time commutators  $[\cdots[\hat{H},S^{+}(x^{(\zeta)}_{\rho_{1}})],\cdots,S^{+}(x^{(\zeta)}_{\rho_{q-1}})],
S^{+}(x^{(\zeta)}_{\rho_{q}})]$ vanish when $q\geq3$. Namely, one only needs to calculate single and double commutators
of $\hat{H}$ with the operators  $S^{+}(x^{(\zeta)}_{\rho})$.
Since we use the pairing operator (\ref{2}) to construct the eigen-vectors (\ref{6}),
the commutator of the one-body mean-field term of (\ref{H}) with $S^{+}(x^{(\zeta)}_{\rho})$
is given by (\ref{3}), while
(\ref{4}) {can} be expressed in terms of the collective operators $S^{+}(x)$ and $S^{+}$
appearing on the right-hand-side of (\ref{3}) when the commutator is applied to the vacuum state $\vert JM\rangle$ with

\begin{eqnarray}\label{7}
[\hat{H}_{\rm P},~S^{+}(x)]\vert JM\rangle=
\sum_{t',t}g_{{t'},{t}}\,S^{+}_{j_{t'}}{2S_{j_{t}}^{0}\over{2\epsilon_{{t}}-x}}\vert JM\rangle=
(\alpha(x)\, S^{+}+\beta(x)\, S^{+}(x))\vert JM\rangle.
\end{eqnarray}
After solving the above binomial equations of the local operators $S^{+}_{j_{1}}$ and $S^{+}_{j_{2}}$, one obtains

\begin{eqnarray}\label{9}\nonumber
&\alpha(x)=-{(x-2\epsilon_{{2}})\left((x-2\epsilon_{{1}})g_{1,1}-(x-2\epsilon_{{2}})g_{1,2}\right)
(\Omega_{{1}}-\nu_{{1}})+
(x-2\epsilon_{{1}})\left((x-2\epsilon_{{1}})g_{1,2}-(x-2\epsilon_{{2}})g_{2,2}\right)(\Omega_{{2}}-\nu_{{2}})
\over{
2(x-2\epsilon_{{1}})(x-2\epsilon_{{2}})(\epsilon_{{1}}-\epsilon_{{2}})}},\\
&\beta(x)=-{(x-2\epsilon_{{2}})(g_{1,1}-g_{1,2})(\Omega_{{1}}-\nu_{{1}})+
(x-2\epsilon_{{1}})(g_{1,2}-g_{2,2})(\Omega_{{2}}-\nu_{{2}})\over{
2(\epsilon_{{1}}-\epsilon_{{2}})}},
\end{eqnarray}
where
the condition $g_{2,1}=g_{1,2}$ is used. It is obvious that (\ref{9}) also assumes
$\epsilon_{{1}}\neq\epsilon_{{2}}$, which is valid for non-degenerate cases.
It is clear that the expression shown on the right-hand-side of (\ref{7})
is impossible when the number of orbits $p\geq3$.
For the standard pairing interaction with $g_{t,t'}=G$ $\forall~t,t'$, (\ref{7})
becomes the commutators shown in Richardson's work~\cite{Ri,duk} with $\beta(x)=0$.
Similarly, the double commutator $S^{+}(x,y)=[[\hat{H}_{\rm P},~S^{+}(x)], ~S^{+}(y)]$ given in (\ref{5})
 is
a  homogenous binomial of degree $2$ in $S^{+}_{j_{t}}$ with $t=1,2$ for the two $j$-orbit case,
which, therefore, can be expressed in terms of $3$ independent terms.
Hence, similar to the commutators shown in the SPM,
one {can} write (\ref{5}) as
\begin{equation}\label{10}
S^{+}(x,y)=
2\sum_{t',t}g_{{t'},{t}}\,{1\over{(2\epsilon_{{t}}-x)
(2\epsilon_{{t}}-y)}}S^{+}_{j_{t'}}S_{j_{t}}^{+}=a(x,y)\, S^{+}S^{+}(x)+b(x,y)\,S^{+}S^{+}(y)+
c(x,y)\,S^{+}(x)S^{+}(y),
\end{equation}
which expressed in terms of $S^{+}$, $S^{+}(x)$, and $S^{+}(y)$ is only possible for
two $j$-orbit case. For a system with $p$ $j$-orbits, $p(p+1)/2$ terms are needed on the right-hand-side of (\ref{10}).
For example,
six terms on the right-hand-side of (\ref{10}) are needed
for {the} three $j$-orbit case. Hence, though it is possible to solve a multi $j$-orbit
system  by using this procedure,  the results
will be very complicated with $p$ variables for a two-pair state.
After comparing the coefficients
of  $S^{+}_{j_{t}}S^{+}_{j_{t'}}$ with the same $t$ and $t'$
on both sides of (\ref{10}), one gets
\begin{eqnarray}\label{12}\nonumber
&a(x,y)={1\over{2(x-y)(\epsilon_{{1}}-\epsilon_{{2}})^{2}}}F(x,y),~~b(x,y)=a(y,x),\\ \nonumber
&c(x,y)={1\over{2(\epsilon_{{1}}-\epsilon_{{2}})^{2}}}\left(
(x+y)(2\epsilon_{{2}}(g_{1,2}-g_{1,1})+2\epsilon_{{1}}(g_{1,2}-g_{2,2}))\right.+\\
&\left. x\,y\,(g_{1,1}+g_{2,2}-2g_{1,2})+4\epsilon_{{2}}^{2}(g_{1,1}-g_{1,2})+4\epsilon_{{1}}^{2}(g_{2,2}-g_{1,2})\right),
\end{eqnarray}
where
\begin{eqnarray}\label{13}\nonumber
&F(x,y)=x\,( 2\epsilon_{{1}}(g_{1,1}-g_{1,2})+2\epsilon_{{2}}(g_{2,2}-g_{1,2}))+
y\,(2\epsilon_{{2}}(g_{1,1}-g_{1,2})+2\epsilon_{{1}}(g_{2,2}-g_{1,2}))+\\
&x\,y\,(2g_{1,2}-g_{1,1}-g_{2,2})+
4g_{1,2}(\epsilon_{{1}}^{2}+\epsilon_{{2}}^{2})-4\epsilon_{{1}}\epsilon_{{2}}
 (g_{1,1}+g_{2,2}),
\end{eqnarray}
and $c(x,y)$ is obviously symmetric in $x$ and $y$.

Using Eqs. (\ref{3}) ,(\ref{7}), and (\ref{10}), one can directly check that
\begin{eqnarray}\label{14}
&\sum_{t}\epsilon_{t}\,\hat{N}_{j_{t}}\vert \zeta,~k;JM\rangle=
\sum_{i}^{k}S^{+}\prod_{\rho\,(\neq i)}^{k}S^{+}(x^{(\zeta)}_{\rho}) \vert JM\rangle+
\sum_{i}^{k}x_{i}^{(\zeta)}\prod_{\rho}^{k}S^{+}(x^{(\zeta)}_{\rho}) \vert JM\rangle
\end{eqnarray}
and
\begin{eqnarray}\label{15}\nonumber
&\hat{H}_{\rm P}\vert \zeta,~k;JM\rangle=\sum^{k}_{i}
\alpha(x^{(\zeta)}_{i})\,S^{+}\prod_{\rho\,(\neq i)}^{k}S^{+}(x^{(\zeta)}_{\rho}) \vert JM\rangle
+\sum^{k}_{i}\beta(x^{(\zeta)}_{i})\prod_{\rho}^{k}S^{+}(x^{(\zeta)}_{\rho}) \vert JM\rangle
+\\
&\sum_{i}^{k}\sum^{k}_{i'\,(\neq i)}a(x_{i'}^{(\zeta)},x_{i}^{(\zeta)} )\,S^{+}\prod_{\rho\,(\neq i)}^{k}S^{+}(x^{(\zeta)}_{\rho})
\vert JM\rangle+
\sum^{k}_{i}\sum^{k}_{i'=i+1}c(x_{i}^{(\zeta)},x_{i'}^{(\zeta)} )\prod_{\rho}^{k}S^{+}(x^{(\zeta)}_{\rho})
\vert JM\rangle.
\end{eqnarray}
With (\ref{14}) and (\ref{15}),
one can prove that the eigen-equation $\hat{H}\vert \zeta, k;JM\rangle=E^{(\zeta)}_{k}\vert \zeta,k;JM\rangle$
is fulfilled if and only if
\begin{equation}\label{12}
1+\alpha(x^{(\zeta)}_{i})+\sum^{k}_{i'\,(\neq i)}a(x_{i'}^{(\zeta)},x_{i}^{(\zeta)} )=0
~~{\rm for}~~i=1,2,\cdots k,
\end{equation}
with the corresponding eigen-energy
\begin{eqnarray}\label{17}\nonumber
&E^{(\zeta)}_{k}=\sum_{t=1}^{p}\epsilon_{{t}}\nu_{{t}}+
\sum_{i=1}^{k}\left(x_{i}^{(\zeta)}+
\beta(x^{(\zeta)}_{i})+\sum^{k}_{i'=i+1}c(x_{i}^{(\zeta)},x_{i'}^{(\zeta)} )\right)=\\\nonumber
&
\sum_{t=1}^{p}\epsilon_{{t}}\nu_{{t}}+\left(
{(g_{1,1}-g_{1,2})(\Omega_{{1}}-\nu_{{1}})\epsilon_{2}\over{\epsilon_{1}-\epsilon_{2}}}+
{(g_{1,2}-g_{2,2})(\Omega_{{2}}-\nu_{{2}})\epsilon_{1}\over{\epsilon_{1}-\epsilon_{2}}}
+{\epsilon_{1}^2(g_{2,2}-g_{1,2})+\epsilon_{2}^2(g_{1,1}-g_{1,2})\over{
(\epsilon_{1}-\epsilon_{2})^{2}}}(k-1)\right)k+\\\nonumber
&\left(
1-{(g_{1,1}-g_{1,2})(\Omega_{{1}}-\nu_{{1}})\over{2\epsilon_{1}-2\epsilon_{2}}}-
{(g_{1,2}-g_{2,2})(\Omega_{{2}}-\nu_{{2}})\over{2\epsilon_{1}-2\epsilon_{2}}}
+{\epsilon_{2}(g_{1,2}-g_{1,1})+\epsilon_{1}(g_{1,2}-g_{2,2})\over{
(\epsilon_{1}-\epsilon_{2})^{2}}}(k-1)\right)\sum_{i=1}^{k}x_{i}^{(\zeta)}+\\
&{g_{1,1}+g_{2,2}-2g_{1,2}\over{4(\epsilon_{1}-\epsilon_{2})^{2}}}\left(
(\sum_{i=1}^{k}x_{i}^{(\zeta)})^{2}-\sum_{i=1}^{k}(x_{i}^{(\zeta)})^{2}\right),
\end{eqnarray}
where $\sum_{t=1}^{p}\epsilon_{{t}}\nu_{{t}}$ is contributed
from particles in the pairing vacuum.
One can easily check that, when  $g_{t,t'}=G$ $\forall~t,t'$, $\alpha(x)=-G\sum_{t}(\Omega_{{t}}-\nu_{{t}})/({2\epsilon_{{t}}-x})$,
$a(x,y)=2G{/{(x-y)}}$, $\beta(x)=c(x,y)=0$,
with which (\ref{12}) and (\ref{17}) become the
Bethe ansatz equations and the corresponding eigen-energy
of the SPM with $E^{(\zeta)}_{k}=\sum_{i=1}^{k}x_{i}^{(\zeta)}$
known previously~\cite{Ri,duk}. Thus, the solution provided by (\ref{6}), (\ref{12}), and (\ref{17})
include the standard and separable pairing models with two non-degenerate
$j$-orbits as special cases, though the form of the
eigenstates shown in (\ref{6}) for the separable
pairing case with $g_{t,t'}=g_{t}g_{t'}$, where $g_{t}$ ($t=1,\cdots,p$)
is a set of real parameters, looks quite different from that used previously~\cite{ba,Rom,claeys,pan4}.
It should be pointed out that (\ref{12}) also implies $g_{1,2}\neq0$.
There will no solution of (\ref{12}) {when} $g_{1,2}=0$.
Actually, similar to the case with no pairing interaction,
a product of the single-particle states is an eigen-state
of (\ref{H}) when $g_{1,2}=0$.
Hence, $g_{1,2}\neq0$ is assumed.


According to the Heine-Stieltjes correspondence \cite{pan0,guan1},
zeros $\{x_{i}^{(\zeta)}\}$
of the extended Heine-Stieltjes polynomials $y_{k}(x)$
of degree $k$ are roots of
 Eq. (\ref{12}), which should satisfy
the following second-order Fuchsian equation:
\begin{equation}\label{18}
A(x)y_{k}^{\prime\prime}(x)+B(x,k)y^{\prime}_{k}(x)-V(x,k)y_{k}(x)=0.
\end{equation}
Here,
\begin{equation}\label{19}
A(x)={1\over{2}}(x^{2}F_{12}+x\,(F_{1}+F_{2})+F_{0})\prod_{t=1}^{2}(2\epsilon_{{t}}-x)
\end{equation}
is a polynomial of degree $4$, in which
\begin{eqnarray}\label{20}
&F_{1}={\epsilon_{1}(g_{1,1}-g_{1,2})+\epsilon_{2}(g_{2,2}-g_{1,2})\over{(\epsilon_{1}-\epsilon_{2})^{2}}},
~F_{2}={\epsilon_{2}(g_{1,1}-g_{1,2})+\epsilon_{1}(g_{2,2}-g_{1,2})\over{(\epsilon_{1}-\epsilon_{2})^{2}}},
~F_{12}={2g_{1,2}-g_{1,1}-g_{2,2}\over{2(\epsilon_{1}-\epsilon_{2})^{2}}},
~F_{0}={2g_{1,2}(\epsilon_{1}^{2}+\epsilon_{2}^{2}) -2\epsilon_{1}\epsilon_{2}(g_{1,1}+g_{2,2})\over{(\epsilon_{1}-\epsilon_{2})^{2}}},
\end{eqnarray}
the polynomial  $B(x,k)$  of degree $3$ is given as
\begin{equation}\label{21}
B(x,k)/A(x)={2\over{x^{2}F_{12}+x\,(F_{1}+F_{2})+F_{0}}}\left(
\sum_{t=1}^{2} {\alpha_{t}^{(1)}+ \alpha_{t}^{(2)}\,x\over{2\epsilon_{{t}}-x}}-(F_{1}+F_{12}\,x)\,(k-1)-1\right),
\end{equation}
where
\begin{eqnarray}\label{22}\nonumber
&\alpha_{1}^{(1)}={(\epsilon_{1}g_{1,1}-\epsilon_{2}g_{1,2})(\Omega_{{1}}-\nu_{{1}})
\over{\epsilon_{1}-\epsilon_{2}}},
~\alpha_{1}^{(2)}={(g_{1,2}-g_{1,1})(\Omega_{{1}}-\nu_{{1}})
\over{2\epsilon_{1}-2\epsilon_{2}}},\\
&\alpha_{2}^{(1)}={(\epsilon_{1}g_{1,2}-\epsilon_{2}g_{2,2})(\Omega_{{2}}-\nu_{{2}})
\over{\epsilon_{1}-\epsilon_{2}}},
~\alpha_{2}^{(2)}={(g_{2,2}-g_{1,2})(\Omega_{{2}}-\nu_{{2}})
\over{2\epsilon_{1}-2\epsilon_{2}}},
\end{eqnarray}
and $V(x)$ is a Van Vleck polynomial of degree $2$, which is determined according to Eq. (\ref{18}).
Therefore, the polynomial approach for the SPM proposed in  \cite{guan1, guan2}
applies to the this case as well. For given the number of pairs $k$, $k$ zeros $\{x_{i}^{(\zeta)}\}$
of $y(x)$ provides
a solution of (\ref{12}) with the corresponding eigen-energy given by (\ref{17}).

\begin{figure}[H]
 \begin{center}
 \includegraphics[scale=0.68]{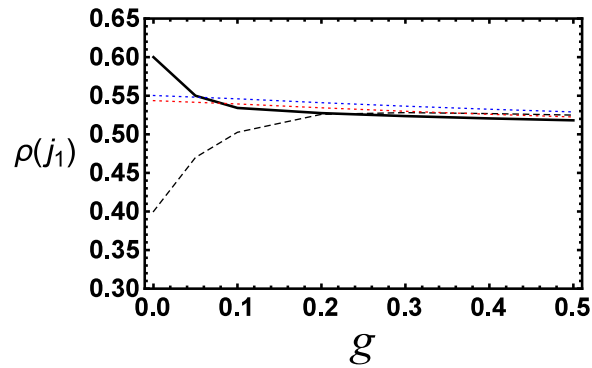}
  \caption{(Color online) The occupation probability of nucleon pairs in the $j_{1}$-orbit
  at the $\zeta$-th excited state for $\zeta=1$ and $\zeta=2$ as a function of $g_{12}=g$ (in MeV)
  with other model parameters the same as those shown in the caption of Table \ref{t1},
where the solid curve represents the  occupation probability at the ground state ($\zeta=1$)
of the NSPM,
the dashed curve is that of the first excited state ($\zeta=2$) of the NSPM, and
the dotted lines from bottom (Red) to the top (Blue) are that of the ground-state and
the first excited state, respectively, in the SPM.} \label{f1}
 \end{center}
 \end{figure}

\vskip .3cm
\noindent {\bf 3. A simple analysis of the model:}~
To demonstrate the use of the solution,
the validity of the SPM is analyzed, of which
only one overall pairing interaction strength can be adjusted.
We consider $5$ pairs in the NSPM with $\epsilon_{1}=1$~MeV and
$\epsilon_{2}=2$~MeV,
$j_{1}=19/2$ and $j_{2}=21/2$, with which each orbit
can accommodate $5$ pairs. The on-site pairing interaction parameters
$g_{1,1}=g_{2,2}=1$~MeV are fixed. We calculated the pair excitation
energies of the NSPM for serval values of $g_{1,2}={g}$, which are presented
in Table \ref{t1}. Then, the overall pairing interaction strength
of the SPM is adjusted according to the
ground-state energy of the NSPM for each case.
Though pairing excitation energies of the SPM are about $2$~MeV different from the  corresponding ones
of the  NSPM,
as shown in Table \ref{t1},
the overlap-square
of the NSPM with the corresponding one of the SPM, $\eta(\zeta)=\vert\langle\zeta\vert\zeta\rangle_{\rm SP}\vert^2$,
is always greater than $94\%$ calculated in this way, where $\vert\zeta\rangle\equiv\vert\zeta,k=5;\,0\,0\rangle$ is obtained
according to (\ref{6}) for each case, while $\vert\zeta\rangle_{\rm SP}$ is the corresponding
eigen-state of the SPM. The results of the overlaps show that the SPM seem
a good approximation to the NSPM.
In fact, with the increasing of the pairing interaction strength $g$ of nucleon pairs
from different orbits, the system undergoes a phase crossover
from localized normal phase mainly determined by the pure mean-field and the on-site pairing interaction
strengths $g_{t,t}$ ($t=1,2$) among nucleon pairs within the same orbits
to the delocalized superconducting  phase, for which there are
a few effective order parameters.
Here we calculate the occupation probability of nucleon pairs in the $j_{1}$
orbit at the $\zeta$-th excited state  defined by
\begin{equation}\label{23}
\rho(j_{1},\zeta)={1\over{k}}\langle\zeta\vert S_{j_{1}}^{+}\frac{\partial}{\partial S_{j_{1}}^{+}}
\vert\zeta\rangle\end{equation}
for $\zeta=1$ and $\zeta=2$.
As clearly shown in Fig. \ref{f1}, the ground-state (the first excited state)
occupation probability of the NSPM decreases (increases) with the increasing of $g$
noticeably around $g\sim0.05${--}0.1~MeV, and there is a crossing point around $g\sim0.21$~MeV.
However, the occupation probability of the ground-state is always a little smaller than that
of the first excited state in the SPM, which is opposite to the result of the NSPM when $g$ is smaller
than the value of the crossing point. They gradually decrease with the increasing of $g$
with the overall pairing interaction strength fitted to the ground-state energy of the NSPM, and
become close to those of the NSPM in the strong $g$ limit.
Therefore, the SPM is a good approximation to the NSPM only when the pairing interaction
among nucleon pairs in different orbits is sufficiently strong.
Nevertheless, the SPM {cannot} account for
the actual quantum phase crossover when the pairing interaction
strengths of different orbits are relatively weaker
and differ from those of the same orbits as required, for example,
in the $ds$- and $fp$-shell nuclei~\cite{35,36}.
Moreover, the on-site pairing interaction strengths $g_{t,t}$
can also change the actual ordering of the single-particle energies.
For example, when $g_{2,2}$ is sufficiently greater than $g_{1,1}$,
the ground state of the system may be dominated by the nucleon
pairs of the $j_{2}$-orbit though $\epsilon_{2}$ is greater than
$\epsilon_{1}$, which may be used to elucidate the inversion of the
single-particle energy ordering of a shell model. Obviously, these phase transition
associated issues {cannot} be described by the SPM,
for which the NSPM  should be adopted.

\begin{table}[H]
\caption{Excited level energies $E^{(\zeta)}$ (in MeV) of the NSPM and the
overlap-square  $\eta(\zeta)=\vert\langle\zeta\vert\zeta\rangle_{\rm SP}\vert^2$
of the pairing excited states with the corresponding ones of the
SPM for $k=5$ pairs over  $j_{1}=19/2$
and $j_{2}=21/2$ orbits with single-particle energies $\epsilon_{1}=1$~MeV,
$\epsilon_{2}=2$~MeV, and $g_{1,1}=g_{2,2}=1$~MeV, where $g_{1,2}=g$, and $\delta g=g-g_{1,1}$ (in MeV).
The overall pairing strength in the SPM is adjusted to reproduce the same ground-state
energy of the NSPM for each case, with which the
corresponding overlap $\eta(\zeta)$ is obtained.
\label{t1}}
\begin{center}\begin{tabular}{ccccccccccccccccccccc}
\hline \hline
  &&$0^{+}_{1}$  &$0^{+}_{2}$  &$0^{+}_{3}$ &$0^{+}_{4}$ &$0^{+}_{5}$ &$0^{+}_{6}$\\
\hline
$\delta g=-0.50$  &$E^{(\zeta)}$ &$-48.95$ &$-36.66$ &$-26.23$ &$-17.30$ &$-9.90$&$-4.96$\\
&$\eta(\zeta)$ &99.600\% &98.989\% &97.968\% &96.600\% &95.370\% &94.548\%\\
$\delta g=-0.25$ &$E^{(\zeta)}$  &$-59.35$ &$-42.61$ &$-28.02$ &$-15.10$ &$-3.86$&$4.93$\\
&$\eta(\zeta)$ &99.949\%&99.870\%&99.756\%&99.653\%&99.714\%&99.714\%\\
$\delta g$=0.25 &$E^{(\zeta)}$ &$-80.28$ &$-54.81$ &$-31.93$ &$-10.96$ &$8.34$&$25.66$\\
&$\eta(\zeta)$&99.977\%&99.946\%&99.905\%&99.883\%&99.9287\%&99.929\%\\
$\delta g$=0.50 &$E^{(\zeta)}$ &$-90.78$ &$-60.98$ &$-33.94$ &$-8.91$ &$14.49$&$36.12$\\
&$\eta(\zeta)$&99.934\%&99.842\%&99.728\%&99.674\%&99.806\%&99.818\%\\
\hline \hline
\end{tabular}
\end{center}
\end{table}

\noindent {\bf 4. Summary:}
In this work, it is shown that the nuclear spherical mean-field plus orbit-dependent non-separable
pairing model with two non-degenerate $j$-orbits, like the standard and separable pairing models,
is also exactly solvable. The solution of the model by using the Bethe ansatz method is presented.
The extended one-variable Heine-Stieltjes polynomials associated to the Bethe ansatz
equations of the solution are determined.
As the use of the solution, a comparison of the solution 
to that of the standard pairing interaction with constant
interaction strength among pairs in any orbit is made via 
a concrete example. It is shown that the overlaps of eigenstates of the model with
those of the standard pairing model are {always large,} especially
for the ground and the first excited state. However,
the quantum phase crossover in the non-separable pairing model
{cannot} be accounted for by the standard pairing interaction, for which
the NSPM should be adopted.

\vskip .3cm
\noindent{\bf Acknowledgement:}~
{Support from the National Natural Science Foundation of China (11675071, 11747318),
the  U. S. National Science Foundation (OIA-1738287 and ACI -1713690),
{U. S. Department of Energy (DE-SC0005248)}, the Southeastern Universities Research Association,
the China-U. S. Theory Institute for Physics with Exotic Nuclei (CUSTIPEN) (DE-SC0009971),
and the LSU--LNNU joint research
program (9961) is acknowledged.}

\end{document}